# Chaotic multi-objective optimization based design of fractional order $PI^\lambda D^\mu$ controller in AVR system

Indranil Pan[a,b], Saptarshi Das[b]

[a] *Centre for Energy Studies, Indian Institute of Technology Delhi, Hauz Khas, New Delhi 110 016, India*

[b] *Department of Power Engineering, Jadavpur University, Salt Lake Campus, LB-8, Sector 3, Kolkata-700098, India.*

Emails: indranil.jj@student.iitd.ac.in, indranil@pe.jusl.ac.in (I. Pan)

saptarshi@pe.jusl.ac.in (S. Das)

## Abstract

In this paper, a fractional order (FO) $PI^\lambda D^\mu$ controller is designed to take care of various contradictory objective functions for an Automatic Voltage Regulator (AVR) system. An improved evolutionary Non-dominated Sorting Genetic Algorithm II (NSGA II), which is augmented with a chaotic map for greater effectiveness, is used for the multi-objective optimization problem. The Pareto fronts showing the trade-off between different design criteria are obtained for the $PI^\lambda D^\mu$ and PID controller. A comparative analysis is done with respect to the standard PID controller to demonstrate the merits and demerits of the fractional order $PI^\lambda D^\mu$ controller.

## 1. Introduction

Multiple generators in a power station are connected to a common bus bar and each of these generators has an automatic voltage regulator (AVR) whose main objective is to control the primary voltage [1]. Due to system disturbances the electrical oscillations may occur for a long time and might result in system instability. Hence effective control algorithms are required to alleviate these issues.

Various controller tuning approaches have been proposed in literature for AVR systems. Traditional controller tuning methods like the minimum variance and pole placement techniques have been used in designing self-tuning regulators [2]. An expert system approach has been proposed in [3] for supervisory actions on an AVR system. Controller design in power systems, using intelligent stochastic optimization methods has recently received a lot of attention from contemporary researchers [4–7]. Particle Swarm Optimization (PSO) has been implemented for the design of AVR systems in [8]. An Intelligent fuzzy logic controller (FLC) has been used in [9] for load frequency control and the fuzzy parameters have been tuned with a PSO algorithm. In [10] a multi-objective optimization approach has been used to tune a lag-lead compensator for an AVR augmented with a power system stabilizer. A hybrid Genetic Algorithm and Bacterial Foraging Algorithm has been used to tune PID controller in AVR in [11]. A craziness based PSO has been used in [12] to tune the PID parameters for a Power System Stabilizer (PSS) controlled AVR system. In [13] a chaotic Ant swarm based optimization has been proposed for tuning the AVR parameters. It is shown that the chaotic ant swarm based optimization performs better than a Genetic algorithm based optimization approach and the designed PID controllers using the proposed method work better. A PSO based fuzzy logic controller has been designed in [14] for automatic generation and control in a two area restructured power system.

However most of the literatures focus on single objective optimization and try to achieve good results for a particular objective like time domain performance or robustness etc. But in practical control system design there is always a trade-off between these objectives. For example it is not possible for a simple linear controller (like the PID controller) to give the fastest settling time along with the best robustness criteria (i.e. best settling time for all range of operating conditions or parameter mismatches). Hence it is natural to formulate a strategy so that many different solutions on a Pareto frontier can be obtained. Then the designer should then be able to choose the appropriate controller depending on the trade-off between the contradictory objectives and his specific requirement for the problem.

Recently fractional order $PI^\lambda D^\mu$ controller [15] is gaining popularity due to its extra flexibility to meet design specifications. Though till date, improved design of fractional order controllers have been restricted mostly to the process control community [16], in few contemporary literatures, FOPID controllers have found applications in the power system community as well. Zamani *et al.* [17] designed a PSO based FOPID controller where the objective function is weighted summation of various time and frequency domain performance criteria. They have shown that FOPID controller outperforms PID controller for ensuring higher robust stability under modelling uncertainty and not for the set-point tracking performance. Tang *et al.* [18] designed FOPID controller with Chaotic Ant Swarm (CAS) algorithm, where the authors claimed to achieve improved time domain performance with the FOPID controller over PID controllers with or without uncertainty in the AVR system components. Alomoush [19] designed FOPID controllers for load-frequency control and Automatic Generation Control (AGC). Though single objective design of fractional order $PI^\lambda D^\mu$ controllers have already been investigated in various control application, its multi-objective design has not yet been so popular. Very few literatures focussed on multi-objective design of $PI^\lambda D^\mu$ controllers but not on its comparison with classical PID e.g. frequency domain design using NSGA-II by Meng and Xue [20] and time domain design using SPEA by Tehrani *et al.* [21]. To the best of the author's knowledge this is the first paper to investigate the multi-objective optimization framework for the fractional order PID controller and make a comparative analysis with the traditional PID controller with respect to the AVR application.

The rest of the paper is organised as follows. Section 2 describes the linearized model of the AVR system. Section 3 describes the basics of the fractional calculus with the different definitions of the fractional differ-integrals. Section 4 describes the fractional order PI^λD^μ controller structure, various sets of contradictory objective functions for optimization and the chaotic multi objective NSGA II algorithm. This is followed by the results and discussions in Section 5 followed by the conclusions and the references.

## 2. Description of the AVR system

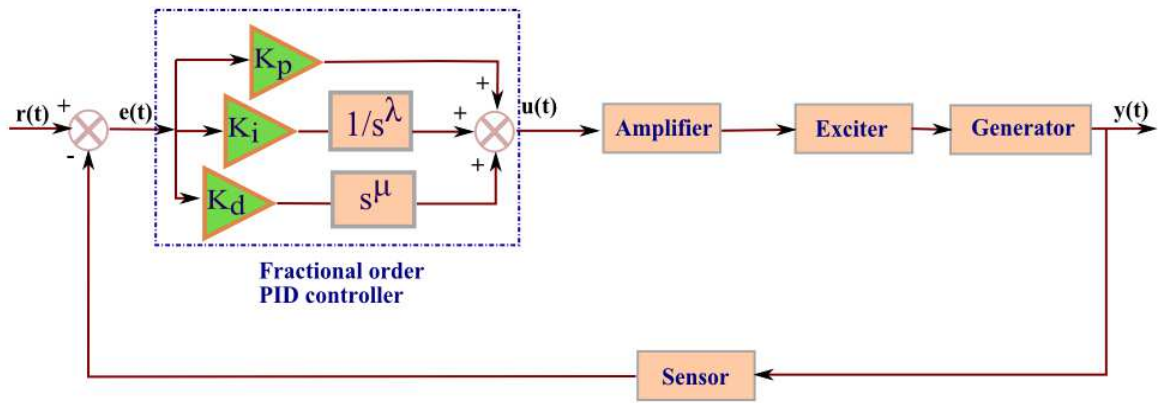

**Figure 1: Linearized model of the AVR system with the Fractional order PID controller**

The AVR model consists of the amplifier model, exciter model, generator model and the sensor model. The representative schematic diagram of the system in shown in Figure 1. The representative transfer functions for these systems along with their range of values are given next [8]:

a) Amplifier model: $\dfrac{K_A}{1+\tau_A s}$

where $10 < K_A < 400$ and a small time constant is in the range $0.02 < \tau_A < 0.1$

b) Exciter model: $\dfrac{K_E}{1+\tau_E s}$

where $10 < K_E < 400$ and a time constant is in the range $0.5 < \tau_E < 1$

c) Generator model: $\dfrac{K_G}{1+\tau_G s}$

where $0.7 < K_G < 1$ and a time constant is in the range $1 < \tau_G < 2$. These constants vary depending on the load.

d) Sensor model: $\dfrac{K_S}{1+\tau_S s}$

where the time constant is in the range $0.001 < \tau_S < 0.06$

The values chosen for this are similar to those in [8]. Thus for the amplifier model $K_A = 10$ and $\tau_A = 0.1$. For the Exciter model $K_E = 1$ and $\tau_E = 0.4$. For the generator model $K_G = 1$ and $\tau_G = 1$. For the sensor model $K_S = 1$ and $\tau_S = 0.01$.

## 3. Basics of Fractional Calculus

The generalized fractional differentiation and integration has mainly three definitions, the Grunwald-Letnikov (G-L) definition and the Riemann-Liouville (R-L) definition and the Caputo definition.

### *3.1. Grunwald-Letnikov (G-L) definition*

This formula is basically an extension of the backward finite difference formula for successive differentiation. This formula is widely used for the numerical solution of fractional differentiation or integration of a function. By Grunwald-Letnikov method the $\alpha^{th}$ order differ-integration of a function $f(t)$ is defined as:

$$D_t^\alpha f(t) := \lim_{h \to 0} \frac{1}{h^\alpha} \sum_{j=0}^{\infty} (-1)^j \binom{\alpha}{j} f(t - jh) \tag{1}$$

where, $\binom{\alpha}{j} = \frac{\alpha!}{j!(\alpha - j)!} = \frac{\Gamma(\alpha+1)}{\Gamma(j+1)\Gamma(\alpha - j + 1)}$ denotes the binomial co-efficients.

The Laplace transform of Grunwald-Letnikov fractional differ-integration is

$$\int_0^\infty e^{-st} {}_0D_t^\alpha f(t) dt = s^\alpha F(s) \tag{2}$$

### *3.2. Riemann-Liouville (R-L) definition*

This definition is an extension of n-fold successive integration and is widely used for analytically finding fractional differ-integrals. By the Riemann-Liouville formula the $\alpha^{th}$ order integration of a function $f(t)$ is defined as:

$$_aI_t^\alpha f(t) = {}_aD_t^{-\alpha} f(t) := \frac{1}{\Gamma(-\alpha)} \int_a^t \frac{f(\tau)}{(t-\tau)^{\alpha+1}} d\tau \tag{3}$$

for $a, \alpha \in \mathbb{R}, \alpha < 0$

By this formula fractional order differentiation is defined as the integer order successive differentiation of a fractional order integral. i.e.

$$_aD_t^\alpha f(t) := \frac{1}{\Gamma(n-\alpha)} \frac{d^n}{dt^n} \int_a^t \frac{f(\tau)}{(t-\tau)^{\alpha-n+1}} d\tau \tag{4}$$

where $n-1 < \alpha < n$

The Laplace transform of Riemann-Liouville fractional differ-integration is:

$$\int_0^\infty e^{-st}{}_0D_t^\alpha f(t)dt = s^\alpha F(s) - \sum_{k=0}^{n-1} s^k {}_0D_t^{\alpha-k-1} f(t)\big|_{t=0} \tag{5}$$

### *3.3. Caputo definition*

In the fractional order systems and control related literatures mostly the Caputo's fractional differentiation formula is referred. This typical definition of fractional derivative is generally used to derive fractional order transfer function models from fractional order ordinary differential equations with zero initial conditions. According to Caputo's definition, the $\alpha^{th}$ order derivative of a function $f(t)$ with respect to time is given by (6) and its Laplace transform can be represented as (7).

$$D^\alpha f(t) = \frac{1}{\Gamma(m-\alpha)} \int_0^t \frac{D^m f(t)}{(t-\tau)^{\alpha+1-m}} d\tau, \quad \alpha \in \mathbb{R}^+, m \in \mathbb{Z}^+ \tag{6}$$

$$m-1 \leq \alpha < m$$

$$\int_0^\infty e^{-st} D^\alpha f(t) dt = s^\alpha F(s) - \sum_{k=0}^{m-1} s^{\alpha-k-1} D^k f(0) \tag{7}$$

where, $\Gamma(\alpha) = \int_0^t e^{-t} t^{\alpha-1} dt$ is the Gamma function and $F(s) := \int_0^\infty e^{-st} f(t) dt$ is the Laplace transform of $f(t)$. This definition is used in the present paper for realizing the fractional integro-differential operators of the FOPID controller.

## 4. Fractional order PI$^\lambda$D$^\mu$ controller structure and its time domain optimization

For control system analysis and design, it is often considered that the initial conditions of FO differential equations are zero to find out the transfer function representation of the linear FO dynamical system. With such an assumption the time domain operator $D^\alpha$ can simply be represented in frequency domain as $s^\alpha$. In this context, a negative sign in the derivative order $(-\alpha)$ essentially implies a fractional integration operator. The FOPID or PI$^\lambda$D$^\mu$ controller is therefore a weighted sum of such operators with extra degrees of freedom for tuning the weights (controller gains) along with the integro-differential order of the operators. The transfer function representation of a FOPID controller [15] is given in (8)

$$C(s) = K_p + \frac{K_i}{s^\lambda} + K_d s^\mu \tag{8}$$

This typical controller structure has five independent tuning knobs i.e. the three controller gains $\{K_p, K_i, K_d\}$ and two fractional order operators $\{\lambda, \mu\}$. For $\lambda = 1$ and $\mu = 1$ the controller structure (8) reduces to the classical PID controller in parallel structure. In order to implement a control law defined by (8) the Oustaloup's band-limited frequency domain rational approximation technique is used in the present paper and also in most of the recent FO control literatures [22]. In fact, the fractional control law with FO differ-integration can also be implemented using the Grunwald-Letnikov definition which is basically a finite difference approximation of fractional derivative with long memory behaviour. But the rationale behind the choice of frequency domain rational approximation of FOPID controller is that it can be easily implemented in real hardware using higher order Infinite Impulse Response (IIR) type analog or digital filters, corresponding to each fractional order differ-integration in the FOPID controller.

On the other hand, the infinite dimensional nature of the fractional order differentiator and integrator in the FOPID controller structure creates hardware implementation issues in industrial application of FOPID controllers. However, few recent research results show that band-limited implementation of FOPID controllers using higher order rational transfer function approximation of the integro-differential operators gives satisfactory performance in industrial automation [23], [24]. The Oustaloup's recursive approximation, which has been used to implement the integro-differential operators in frequency domain is given by the following expression, representing a higher order analog filter.

$$s^\alpha \simeq K \prod_{k=-N}^{N} \frac{s + \omega'_k}{s + \omega_k} \tag{9}$$

where, the poles, zeros, and gain of the filter can be recursively evaluated as:

$$\omega_k = \omega_b \left(\frac{\omega_h}{\omega_b}\right)^{\frac{k+N+\frac{1}{2}(1+\alpha)}{2N+1}}, \omega'_k = \omega_b \left(\frac{\omega_h}{\omega_b}\right)^{\frac{k+N+\frac{1}{2}(1-\alpha)}{2N+1}}, K = \omega_h^\alpha \tag{10}$$

Thus, any signal $f(t)$ can be passed through the filter (9) and the output of the filter can be regarded as an approximation to the fractionally differentiated or integrated signal $D^\alpha f(t)$. In (9)-(10), $\alpha$ is the order of the differ-integration, $(2N+1)$ is the order of the filter and $(\omega_b, \omega_h)$ is the expected fitting range.

Even with the truncation of infinite dimensional natures of FO operators with high order IIR filters, the obtained FOPID controllers are found to outperform classical PID structure in most recent literatures [22], [24], [25]. Thus there is also a trade-off between the complexity of the realization of the FOPID controller and the achievable accuracy. In the present study, $5^{th}$ order Oustaloup's recursive approximation is done for the integro-differential operators within a frequency band of the constant phase elements (CPEs) as $\omega \in \{10^{-2}, 10^{2}\}$ rad/sec.

## 5. Need for multi-objective optimisation and contradictory objective functions

The question of why multi-objective optimisation is required for controller designing problems is enunciated explicitly in [26]. It states that the key concept of the different design paradigms, like the $H_2$, $H_\infty$ or $L_1$ control is that the design objective can be satisfied by minimizing a weighted norm of the closed loop transfer function. However each norm has its own characteristic feature and minimizing that norm ensures that the control system satisfies that criteria well, but it does not say anything about the other design specifications. For example [26], minimising the $H_2$ norm implies good closed loop stabilisation in the presence of disturbances. But controllers designed with the sole consideration of the $H_2$ norm might not possess guaranteed robust stability. The $H_\infty$ or $L_1$ norms on the other hand give closed-loop robust stability. But the former is a frequency domain technique while the latter is a time domain technique and they address design specifications in frequency domain and time domain respectively. It is known that different conflicting specifications like disturbance attenuation, robust stability, good tracking etc. cannot be represented by a single norm. Thus a multi-objective algorithm is essential for assessing the performance limits and analysing the various trade-offs among the disparate design objectives.

Two sets of contradictory objective functions are used to demonstrate the proposed multi objective optimization algorithm for the fractional and the integer order PID controller. The objective functions along with the rationale of including them are given below:

**Case I**

The two contradictory objective functions that are considered in the first case are the Integral of the Time multiplied Squared Error (ITSE) ($J_1$) and the Integral of the Squared Deviation of Controller Output (ISDCO) ($J_2$). The first objective function $J_1$ tries to ensure fast tracking of the desired set-point. The time multiplication term assigns heavy penalty to the errors occurring at later stages and hence ensures faster settling time.

$$J_1 = ITSE_{set-point} = \int_0^\infty t e_{sp}^2(t) dt \qquad (11)$$

The second objective function $J_2$ tries to reduce the error in the control signal as large control signals would require a larger actuator sizing and consequent increase in cost.

Minimizing the performance index involving the control signal is also required as high oscillations or perturbations in the manipulated variable is not desirable [27], [28]. This is because the manipulated variables are physical quantities and this might result in shocks to the system. $J_2$ is given by equation (12) and the term $\Delta u(t)$ represents the change in the absolute value of the control signal.

$$J_2 = ISDCO = \int_0^\infty \Delta u^2(t) dt \tag{12}$$

$J_1$ and $J_2$ are contradictory objectives since to reduce the steady state tracking error or to obtain fast tracking (i.e. to minimize $J_1$), the controller must exert more effort and hence the value of $J_2$ would increase and vice-versa.

**Case II**

The two objective functions considered in the second case are the Integral of Time Multiplied Squared Error (ITSE) for the set-point ($J_1$) and the ITSE for the load disturbance ($J_3$). $J_1$ tries to minimize the tracking error. The squared error term gives more penalty to the error and ensures an even faster settling time. However this might also result in what is termed as derivative kicks. $J_3$ tries to minimize the deviation from the set-point when unpredictable disturbances occur in the process.

$$J_3 = ITSE_{load\_disturbance} = \int_0^\infty t e_{ld}^2(t) dt \tag{13}$$

The tracking performance is equivalent to the H$_2$ norm of the error in the frequency domain. If 2-norm is lesser, then the tracking performance is better and vice-versa. Load disturbance depends on the infinity norm of the sensitivity function in frequency domain. In Zhuang and Atherton [29], different tuning rules have been developed to fulfil these two different objectives separately. In this case, two objectives are simultaneously optimized through the algorithm.

**Case III**

In the third case, all the functions from the previous cases are taken. So the objectives $J_1$, $J_2$ and $J_3$ are imposed together on the system and the controller designing is done with these three contradictory objectives.

# 6. Multi-objective chaotic Non-dominated Sorting Genetic Algorithm-II (chaotic NSGA-II)

A generalized multi-objective optimization framework can be defined as follows:

Minimize $F(x) = (f_1(x), f_2(x), ..., f_m(x))$

$$F(x) = (f_1(x), f_2(x), ..., f_m(x)) \tag{14}$$

such that $x \in \Omega$

where $\Omega$ is the decision space, $\mathbb{R}^m$ is the objective space, and $F: \Omega \to \mathbb{R}^m$ consists of $m$ real valued objective functions.

Let, $u = \{u_1, ..., u_m\}$, $v = \{v_1, ..., v_m\} \in \mathbb{R}^m$ be two vectors. $u$ is said to dominate $v$ if $u_i < v_i$ $\forall i \in \{1, 2, ..., m\}$ and $u \neq v$. A point $x^* \in \Omega$ is called Pareto optimal if $\nexists$ $x \mid x \in \Omega$ such that $F(x)$ dominates $F(x^*)$. The set of all Pareto optimal points, denoted by PS is called the Pareto set. The set of all Pareto objective vectors, $PF = \{F(x) \in \mathbb{R}^m, x \in PS\}$, is called the Pareto Front. This implies that no other feasible objective vector exists which can improve one objective function without simultaneously worsening of some other objective function.

Multi-objective Evolutionary Algorithms (MOEAs) which use non-dominated sorting and sharing, have higher computational complexity. They use a non-elitist approach and require the specification of a sharing parameter. The non-dominated sorting genetic algorithm (NSGA-II) removes these problems and is able to find a better spread of solutions and better convergence near the actual Pareto optimal front [30]. The pseudo code for the NSGA II is as shown below [30], [31].

NSGA II Algorithm

Step 1: generate population $P_0$ randomly

Step 2: set $P_0 = (F_1, F_2, ...) = $ non-dominated-sort $(P_0)$

Step 3: for all $F_i \in P_0$

    crowding-distance-assignment $(F_i)$

Step 4: set t=0

    while (not completed)

        generate child population $Q_t$ from $P_t$

        set $R_t = P_t \cup Q_t$

        set $F = (F_1, F_2, ...) = $ non-dominated-sort $(R_t)$

        set $P_{t+1} = \phi$

        i=1

        while $|P_{t+1}| + |F_i| < N$

            crowding-distance-assignment $(F_i)$

            $P_{t+1} = P_{t+1} \cup F_i$

            i=i+1

        end

        sort $F_i$ on crowding distances

        set $P_{t+1} = P_{t+1} \cup F_i\left[1:(N - |P_{t+1}|)\right]$

        set $t = t + 1$

    end

return $F_1$

Here $N$ represents the number chromosomes in the population i.e. the population size. The NSGA II algorithm converts $M$ different objectives into one fitness measure by composing distinct fronts which are sorted based on the principle of non-domination. In the process of fitness assignment, the solution set not dominated by any other solutions in the population is designated as the first front $F_1$ and the solutions are given the highest fitness value. These solutions are then excluded and the second non dominated front from the remaining population $F_2$ is created and ascribed the second highest fitness. This method is iterated until all the solutions are assigned a fitness value. Crowding distances are the normalized distances between a solution vector and its closest neighbouring solution vectors in each of the fronts. All the constituent elements of the front are assigned crowding distances to be later used for niching. The selection is achieved in tournaments of size 2 according to the following logic.

    a) If the solution vector lies on a lower front than its opponent, then it is selected.

b) If both the solution vectors are on the same front, then the solution with the highest crowding distance wins. This is done to retain the solution vectors in those regions of the front which are scarcely populated.

The population size is taken as 100 and the algorithm is run until the cumulative change in fitness function value is less than the function tolerance of $10^{-4}$ over 100 generations. The crossover fraction is taken as 0.8 and an intermediate crossover scheme is adopted. The mutation fraction as 0.2. For choosing the parent vectors based on their scaled fitness values, the algorithm uses a tournament selection method with a tournament size of 2. This tournament size has been used in previous studies by other researchers and has given good results [32]. The Pareto front population fraction is taken as 0.7. This parameter indicates the fraction of population that the solver tries to limit on the Pareto front. The optimization variables for the fractional order PID controller are the proportional-integral-derivative gains and the differ-integral orders, i.e. $\{K_p, K_i, K_d, \lambda, \mu\}$. For the integer order PID controller the optimization variables are the gains i.e., $\{K_p, K_i, K_d\}$.

The uniformly distributed random number generator is normally used for the crossover and mutation operations in the standard version of the NSGA-II algorithm [30]. However since the strength of evolutionary algorithms lies in the randomness of the crossover and mutation operators, many contemporary researchers have focussed on increasing the efficiency of these algorithms by incorporating different random behaviours through various techniques like stochastic resonance and noise [33], chaotic maps [34] etc. In [35] it has been shown that the performance of these evolutionary algorithms increase if different types of chaotic maps are introduced instead of the uniform random number generator for the crossover and mutation operations. It has also been demonstrated in [35] that, in general, using chaotic systems for the random number generation in the crossover and mutation operations is better than using random numbers generated from a noisy sequence in terms of convergence and effectiveness of the algorithms in finding global minima. In [36] it has been shown that the multi-objective NSGA-II algorithm can be improved by using chaotic maps and gives better result than the original NSGA-II algorithm in terms of convergence and high efficiency in calculation. This is due to the fact that the chaotic process introduces diversity in the solutions. In this paper, we adopt this policy and use a chaotic logistic map to obtain better solutions and convergence characteristics of the NSGA-II algorithm. The logistic map is one of the simplest discrete time dynamical systems exhibiting chaos. The equation for the logistic map is given as follows:

$$x_{k+1} = ax_{k+1}(1 - x_k) \tag{15}$$

In [37], a comparative performance of different chaotic maps (like the Lozi Map, the Chua oscillator, Gauss Map, Sinusoidal iterator, logistic map etc.) with various Lyapunov exponents have been shown and the chaotic logistic map has been shown to work well for standard test-bench problems. The initial condition of the map in Equation (15) has been chosen to be $x_0 = 0.2027$ and the parameter $a = 4$ has been taken similar to that in [37]. The logistic map is implemented in the NSGA-II algorithm as a replacement for the *rand()* function with uniform random number generation. Hence it should produce a value between 0 and 1, every time there is a function call to it. Now choosing initial values like 0, 0.25, 0.5, 0.75, 1, would all result in some constant value after some iterations. Thus there would be no effect of randomness at all, which is not desirable for effective functioning of the evolutionary algorithm. Thus the parameters of the logistic map (i.e. $a$ and $x_0$) must be chosen in such a manner that the solution of the map keeps on oscillating chaotically between $[0,1]$ and does not settle to some constant value or diverge outside the range of $[0,1]$. To choose the parameters effectively, the bifurcation diagram and studies on chaotic properties of the logistic map needs to be consulted which have been extensively documented in a wide array of literatures [38]. The values chosen here, are proved to give chaotic oscillations and aid in finding effective solutions for evolutionary algorithms [37]. For the optimization problem, the limits of $\{K_p, K_i, K_d\}$ are chosen to be $[0,100]$ and the bounds of the differ-integral orders $\{\lambda, \mu\}$ are chosen to be in the range $[0,2]$.

## 7. Results and Discussions

The Pareto frontiers for Case I with two contradictory objective functions $J_1$ and $J_2$ are shown in Figure 2. The simulation is run for a finite time horizon of 10 seconds. Some representative solutions on the Pareto front are reported in Table 1 for both the PID and the FOPID controllers. The two extreme solutions and the median solution on the Pareto front are chosen as representative cases.

**Table 1: Representative solutions on the Pareto front for Case I**

| Controller | Solution | $J_1$ | $J_2$ | $K_p$ | $K_i$ | $K_d$ | $\lambda$ | $\mu$ |
|---|---|---|---|---|---|---|---|---|
| FOPID | A1 | 1.02842 | 1.05130 | 2.05111 | 1.01165 | 0.45682 | 0.70557 | 1.04794 |
| | B1 | 1.08101 | 1.00445 | 0.61716 | 0.68350 | 0.24933 | 0.56892 | 1.01451 |
| | C1 | 1.18466 | 1.00058 | 0.19357 | 0.58262 | 0.17992 | 0.50392 | 1.01767 |
| PID | A2 | 1.02009 | 1.09681 | 2.76662 | 0.49906 | 0.50078 | - | - |
| | B2 | 1.04292 | 1.01311 | 1.09475 | 0.38383 | 0.23125 | - | - |
| | C2 | 1.18004 | 1.00047 | 0.21406 | 0.13621 | 0.00829 | - | - |

As is evident from Figure 2, the Pareto front for the FOPID controller is totally inside the concave portion of the PID controller's front. Thus the PID controller outperforms the FOPID one for all the cases. For higher values of $J_1$, the Pareto front of both the FOPID and the PID almost merges. Thus if a solution is chosen in this region, then there is not much difference in using the PID or the FOPID controller. Hence when these two contradictory objectives are considered in the design framework, the PID controller should be preferred also due to its structural simplicity.

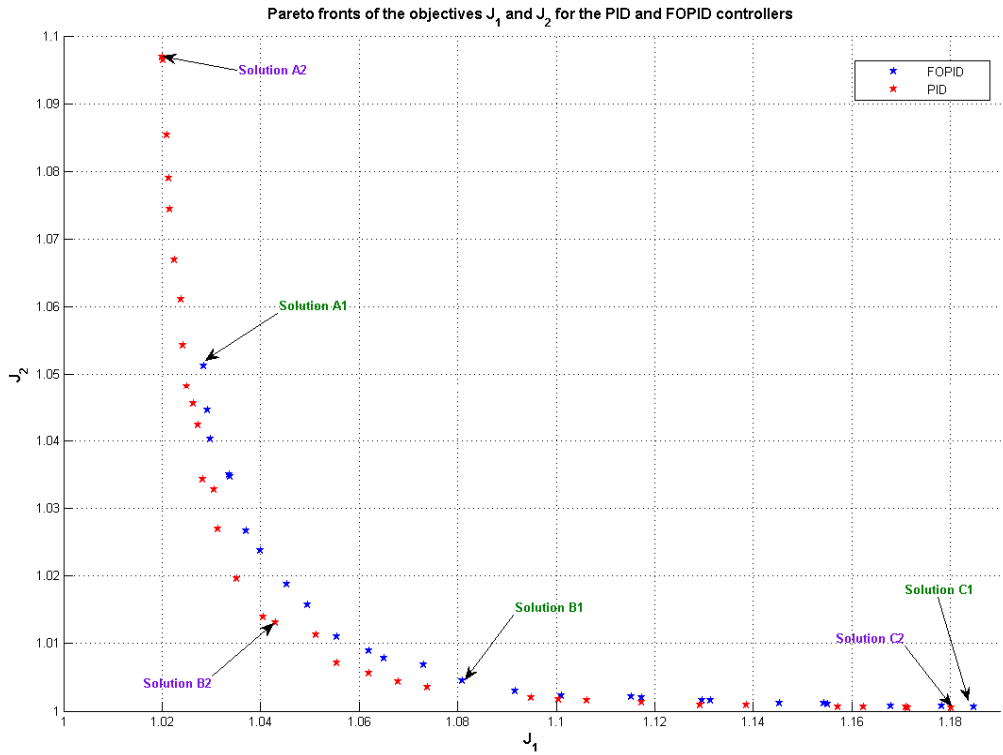

**Figure 2 : Pareto front of the objectives $J_1$ and $J_2$ for the PID and FOPID controllers**

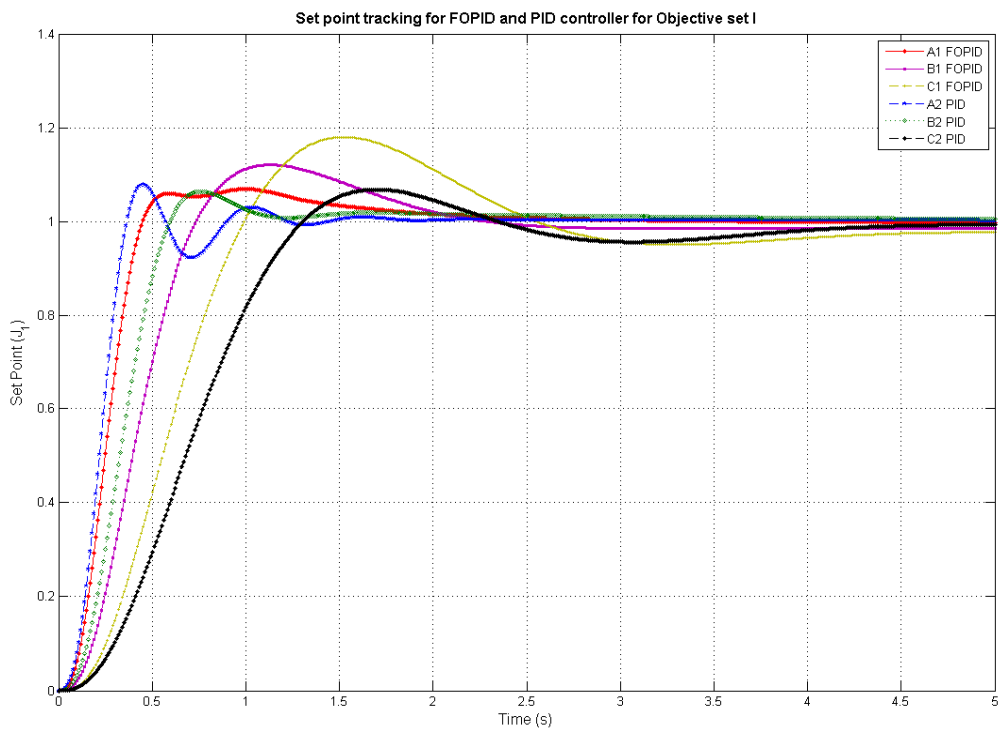

**Figure 3: Set point tracking for representative solutions as reported in Table 1**

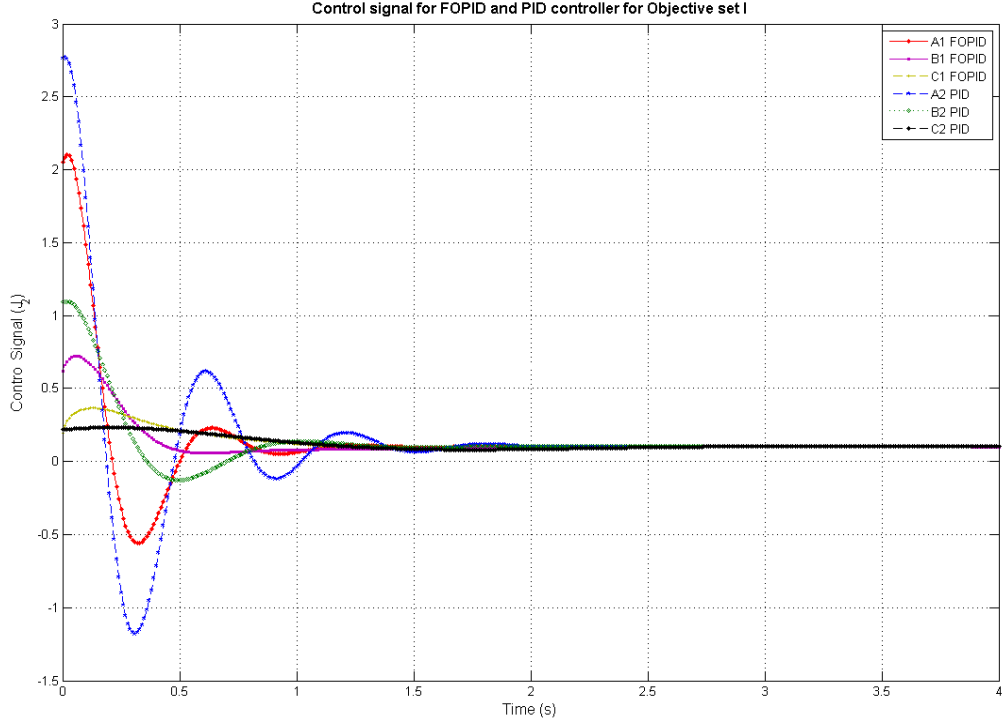

**Figure 4: Control signal for representative solutions as reported in Table 1**

Figure 3 and Figure 4 show the set-point tracking and the control signal respectively, for the representative solutions as reported in Table 1. The result from the Pareto front is also verified by these figures. It can be observed that the solutions of the FOPID controller have a higher overshoot and a larger settling time as compared to the PID controller. From Figure 4 it can be seen that the control signals are higher in the PID case than the FOPID case for lower values of $J_1$, but the PID performs better at higher values of $J_1$.

Here, the Pareto front actually gives the limits of the controller performance. Thus it implies that using a particular controller structure and corresponding conflicting objectives, the designer cannot expect to get any better results. Any result which improves one objective function will have a poor performance measure with respect to the other objective functions. If only a single objective tracking criteria was imposed for set point tracking, then the controller could have tracked the system much better as shown in other studies like [18]. But if using those controller parameters, the load disturbance is checked, it would prove to be much worse than the obtained solutions using this scheme of multi objective optimisation. This is because the multi-objective optimisation produces a set of non-dominated solution [30].

Figure 5 shows the Pareto fronts for the PID and the FOPID controller for case II. Here it can be seen that the FOPID controller outperforms the PID controller for all possible cases as the FOPID Pareto frontier totally encloses the PID one. Table 2 shows some representative solutions on the Pareto fronts for both the PID and the FOPID cases. These representative solutions are chosen as the ones on the extreme ends and the median solution as before.

**Table 2: Representative solutions on the Pareto front for Case II**

| Controller | Solution | $J_1$ | $J_3$ | $K_p$ | $K_i$ | $K_d$ | $\lambda$ | $\mu$ |
|---|---|---|---|---|---|---|---|---|
| FOPID | A3 | 1.00880 | 1.41013 | 1.00338 | 1.32100 | 0.61858 | 0.88479 | 0.99999 |
| | B3 | 1.00943 | 1.19484 | 1.08069 | 2.09803 | 0.69040 | 0.85782 | 0.99999 |
| | C3 | 1.04920 | 1.00328 | 4.58472 | 15.99028 | 1.118637 | 0.99999 | 0.99999 |
| PID | A4 | 1.01398 | 1.79841 | 6.43078 | 0.51842 | 1.13776 | - | - |
| | B4 | 1.01505 | 1.40615 | 6.52483 | 0.78404 | 1.14001 | - | - |
| | C4 | 1.02999 | 1.00350 | 9.12489 | 10.59754 | 1.93362 | - | - |

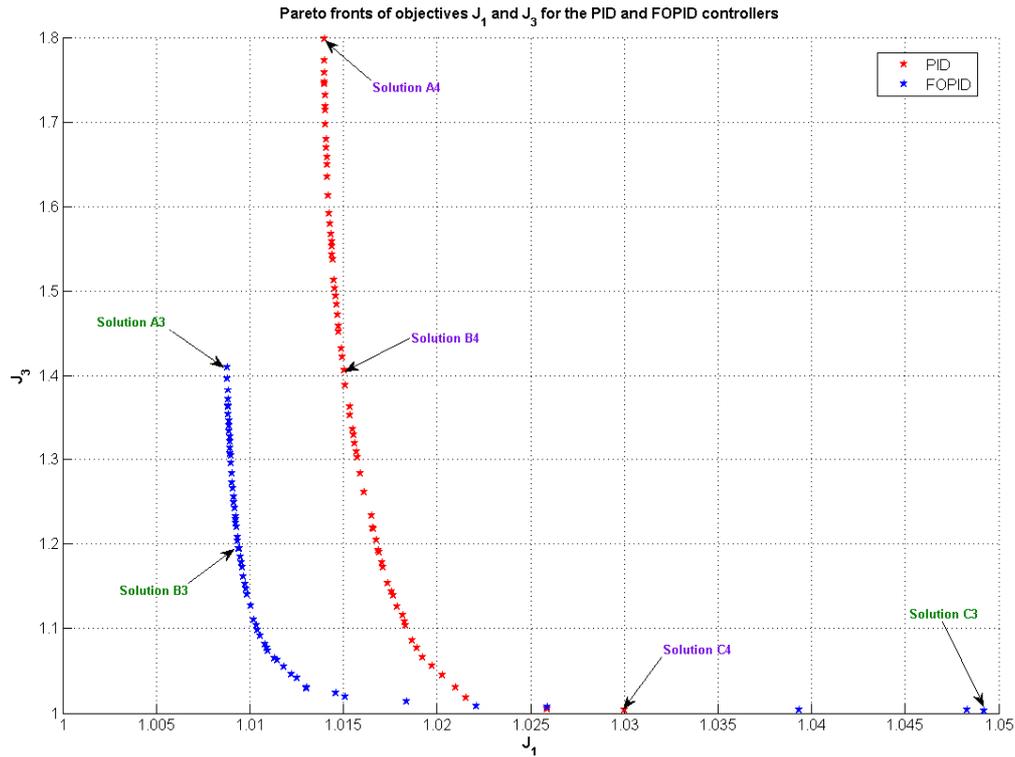

**Figure 5: Pareto fronts of objectives J$_1$ and J$_3$ for the PID and the FOPID controllers**

Figure 6 and Figure 7 show the set point tracking and the load disturbance rejection for the representative solutions as reported in Table 2. Solutions A3 and B3 of the FOPID controller have a faster settling time than the corresponding solutions A4 and B4 of the PID controller. The Solution C3 is more oscillatory than its counterpart Solution C4 as it lies on the far end of the Pareto frontier. The PID Pareto frontier does not extend to such a distance on the right and extends more toward the left. The corresponding load disturbance curves for the PID and the FOPID controllers are shown in Figure 7. The system with the FOPID controller quickly recovers from a unit load disturbance, but the PID controller takes a long time to recover under a load disturbance. Thus the fact that the FOPID controller is better than the PID one for these set of objective functions is validated. This is also evident from the Pareto frontiers themselves in Figure 5. Since the load disturbance rejection of the PID controller does not settle within 10 seconds, thus for this case the simulations are run for a finite time horizon of 20 seconds and the corresponding results are reported.

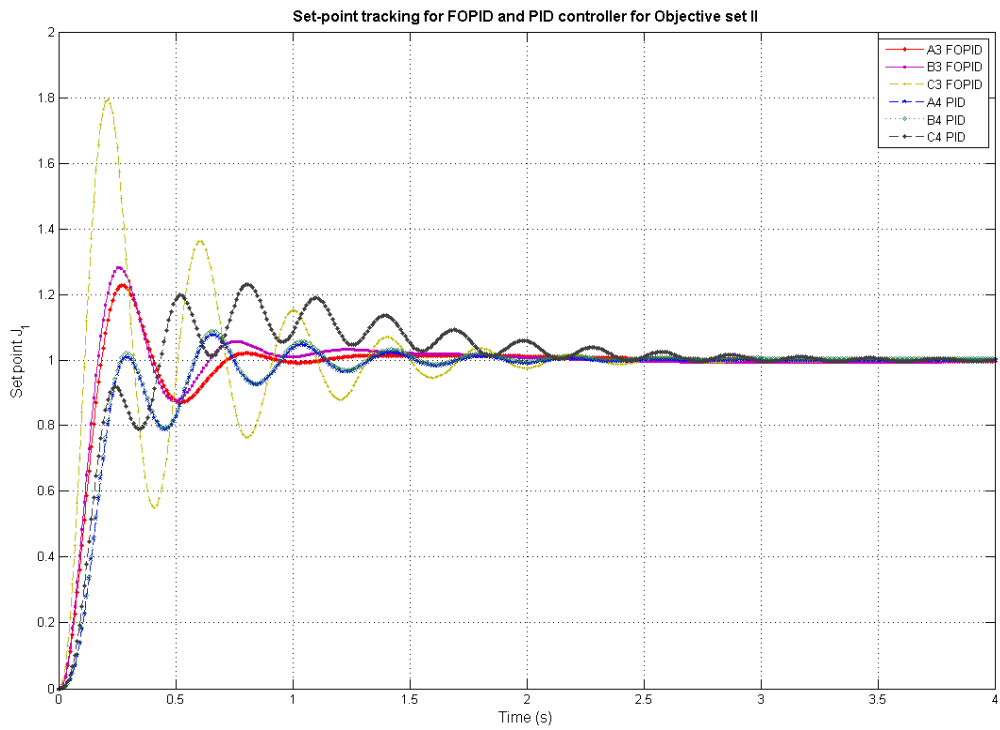

**Figure 6: Set point tracking (Objective $J_1$) for representative solutions as reported in Table 2**

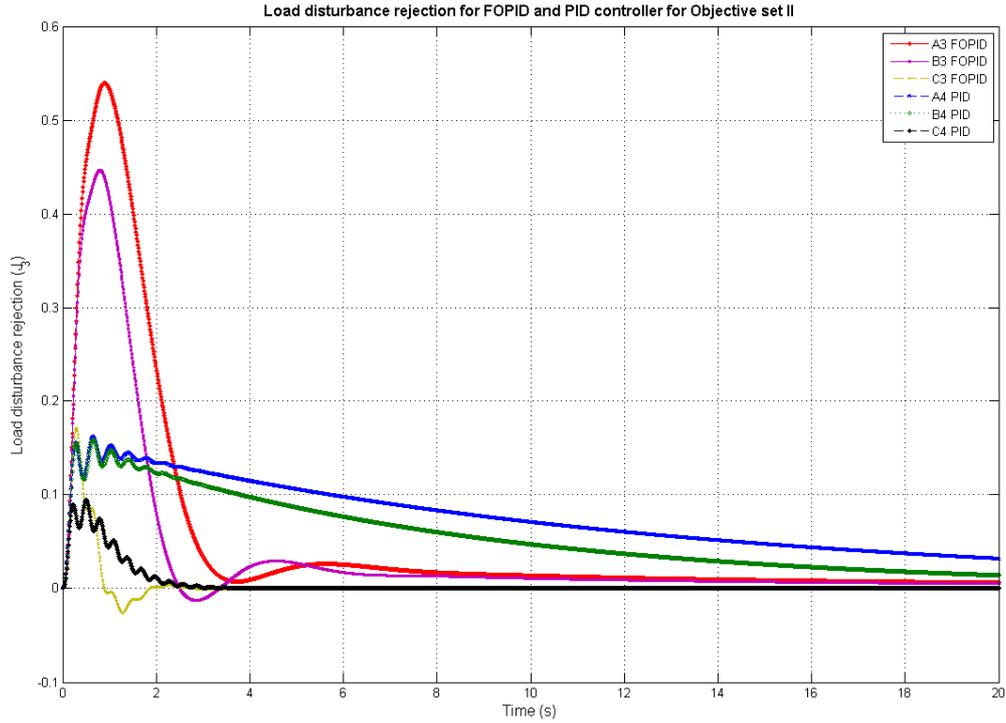

**Figure 7: Load disturbance rejection (Objective $J_3$) for representative solutions as reported in Table 2**

In Case III all the three objective functions are considered together and the multi objective optimization is run. The Pareto frontier obtained in this case is shown in Figure 8. Since discrete solutions are obtained on the Pareto front, hence an interpolated surface is constructed for better visualization of the domain. Some representative solutions from the Pareto front are reported in Table 3.

**Table 3 : Representative solutions on the 3D Pareto front for Case III**

| Controller | $J_1$ | $J_3$ | $J_2$ | $K_p$ | $K_i$ | $K_d$ | $\lambda$ | $\mu$ |
|---|---|---|---|---|---|---|---|---|
| FOPID | 1.00005 | 1.00000 | 2.76967 | 0.98948 | 1.76282 | 0.36743 | 0.94674 | 0.70517 |
|  | 1.1199271 | 1.80696 | 1.11384 | 0.83997 | 1.33590 | 0.35115 | 0.91469 | 0.71071 |
|  | 1.7382979 | 2.52705 | 1.01821 | 0.46675 | 0.95199 | 0.29679 | 0.88723 | 0.23069 |
| PID | 1.0035357 | 1.00000 | 4.93300 | 12.10266 | 6.06725 | 7.70072 | - | - |
|  | 2.7348721 | 2.30651 | 1.00205 | 0.42376 | 2.08440 | 0.54727 | - | - |
|  | 5.0069269 | 2.36431 | 1.00005 | 0.02998 | 1.41050 | 1.24631 | - | - |

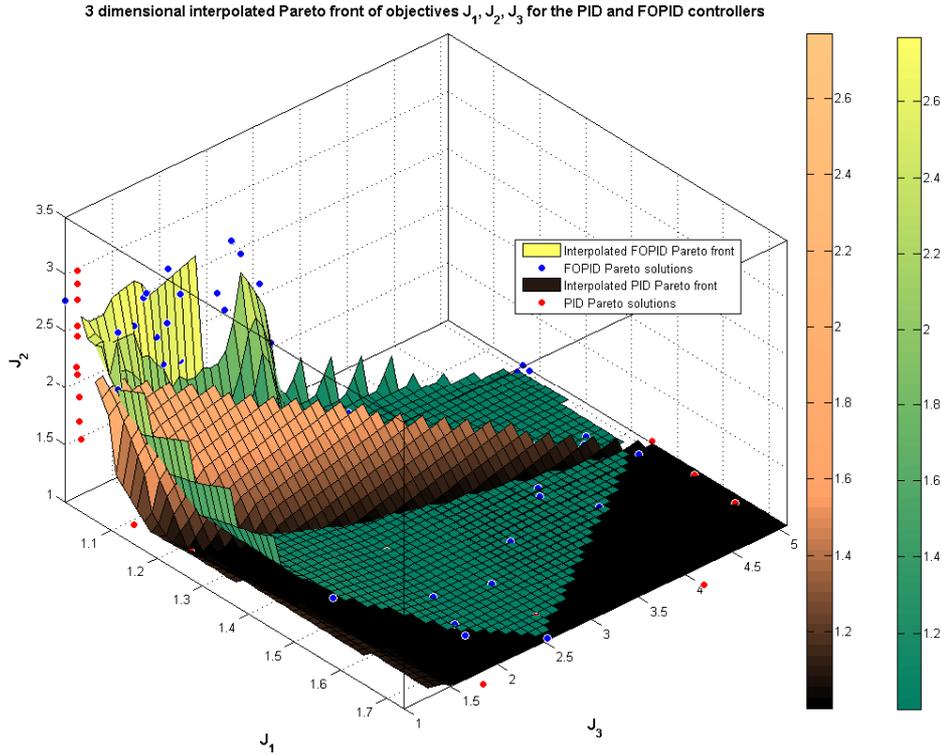

**Figure 8: 3D interpolated Pareto front of objectives J₁, J₂, J₃ for the PID and the FOPID controller**

From Figure 8 it is clear that for these three sets of objective functions, the Pareto fronts of the PID and FOPID controllers have an intersecting region. Thus one single controller does not perform well for all possible cases. Depending on the different weights assigned to the contradictory objective functions by the designer, the FOPID controller will give better performance in some cases and the PID controller will give better performance in other cases. A larger set of representative solutions on the Pareto front for all the three cases is given in the Appendix.

## 8. Robustness analysis of the obtained solutions

The controllers have been designed for the nominal operating conditions. However it is desirable that the tuned controller work satisfactorily for other operating conditions as well, i.e. it must be robust to change in system parameters. To illustrate the effect of the variation in system parameter on the obtained solution, the gain and time constant of the generator ($K_G$ and $\tau_G$) are varied in the limits as specified in Section 2. The variations in the generator transfer function are due to the load changes which occur frequently in the system and the controller must be capable of handling these circumstances. Hence in the present study only this variation is considered for robustness analysis.

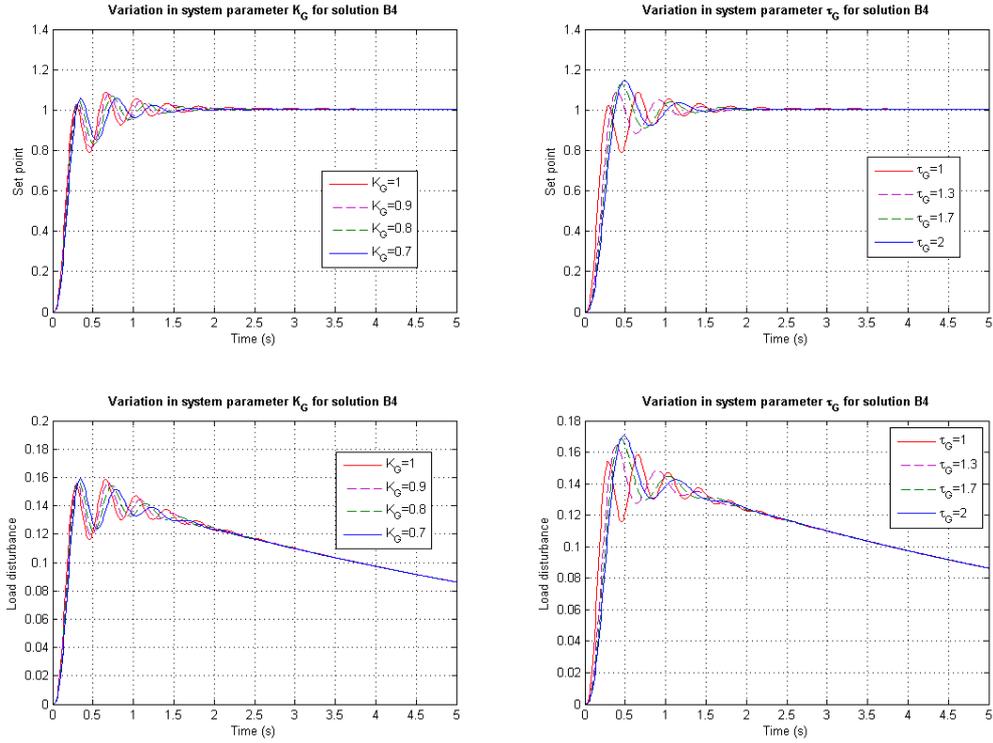

**Figure 9: Robustness analysis of obtained solution B4 for PID controller**

Figure 9 and Figure 10 show the robustness analysis for the PID and the FOPID controller respectively, for the respective median solutions on the Pareto front. It can be seen that both the PID and the FOPID controllers offer sufficient robustness inspite of change in system parameters. From Figure 10, it can be observed that the FOPID gives a much more consistent time domain performance than the PID controller. Hence the FOPID controller is much more capable of tolerating changes in system parameters than the PID controller. However the FOPID has a more complex structure and consequently would be difficult and expensive to implement in actual hardware. Thus the system designer must decide whether the additional gain in performance is significant enough to off-set the higher cost and complexity due to the FOPID controller.

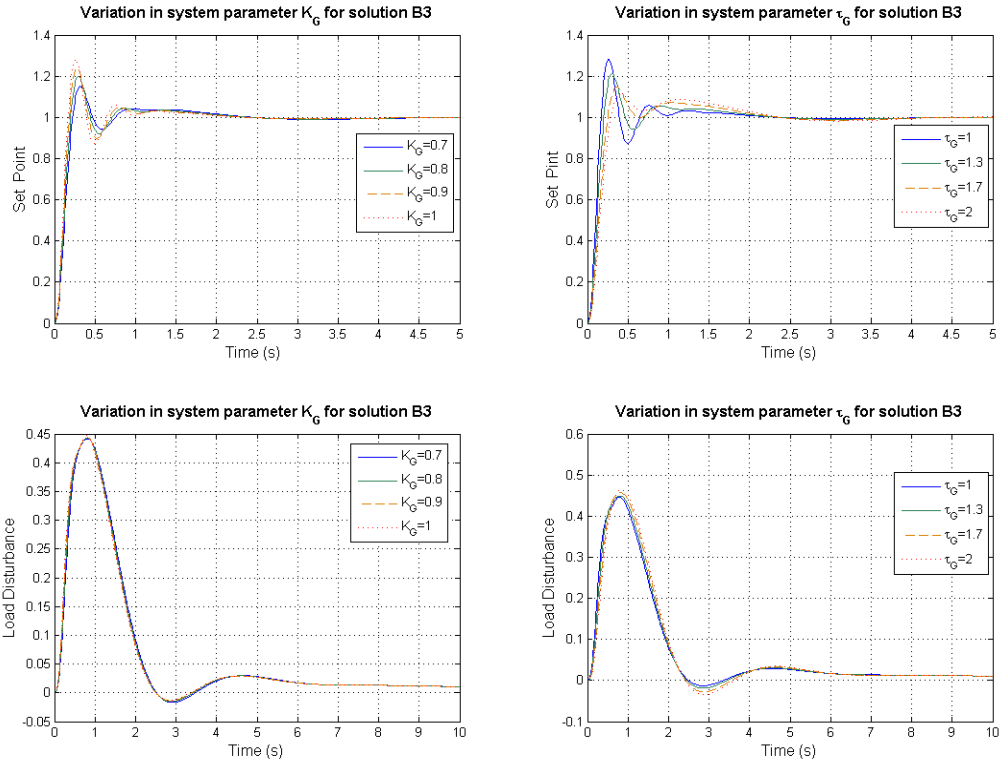

**Figure 10: Robustness analysis of obtained solution B3 for FOPID controller**

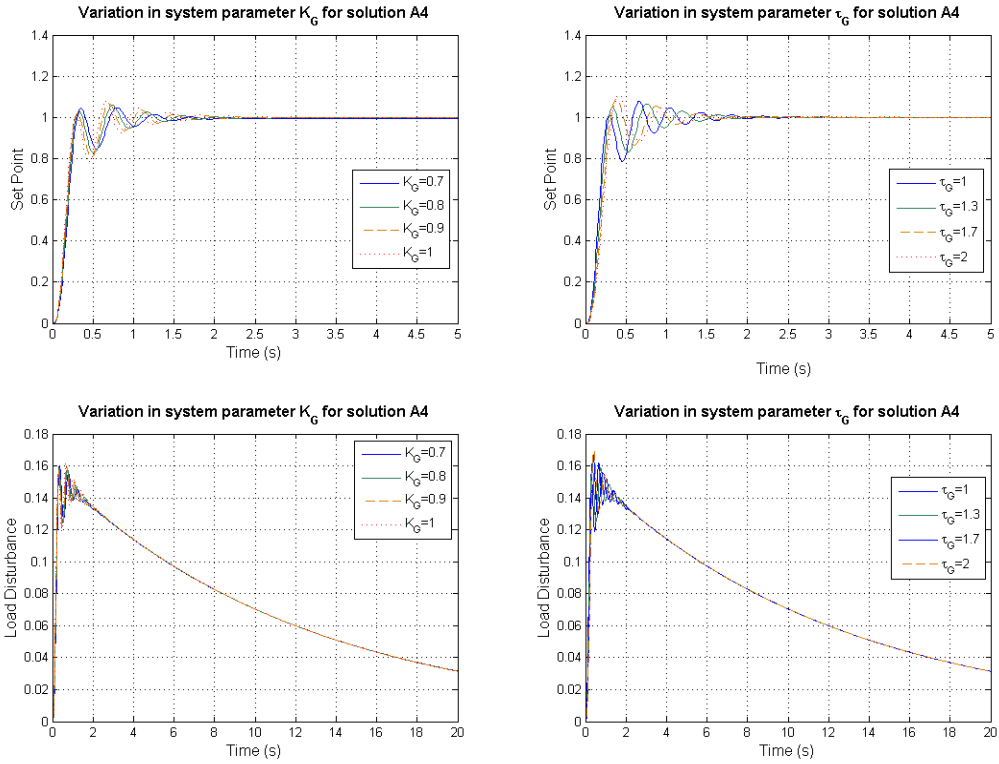

**Figure 11: : Robustness analysis of obtained solution A4 for PID controller**

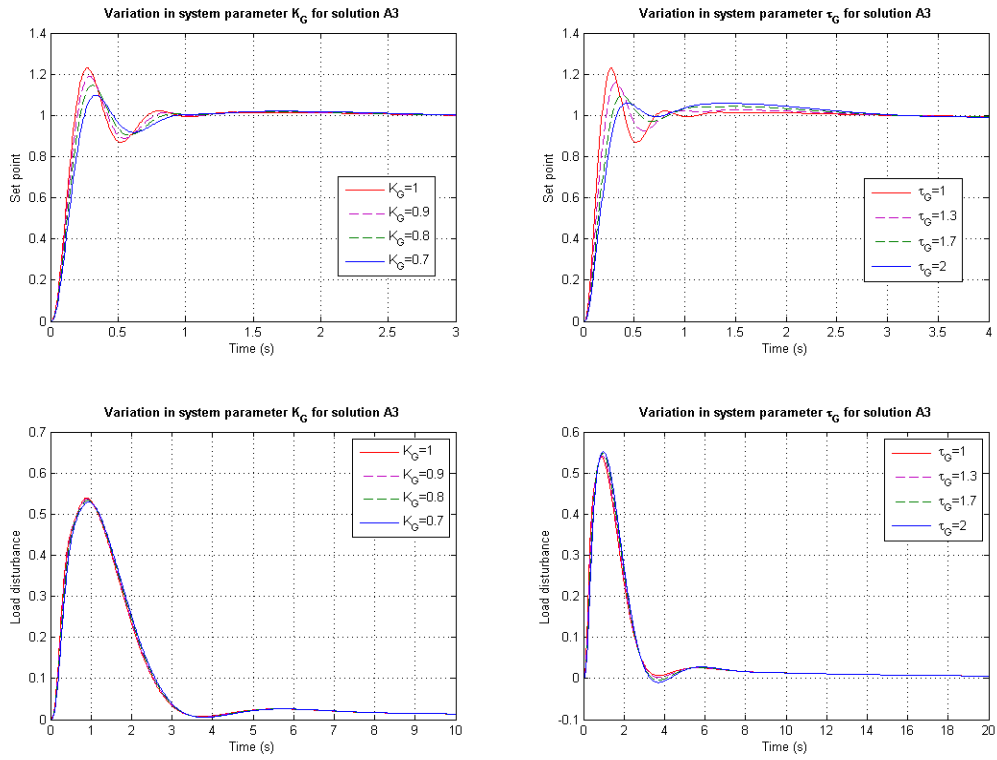

**Figure 12: : Robustness analysis of obtained solution A3 for FOPID controller**

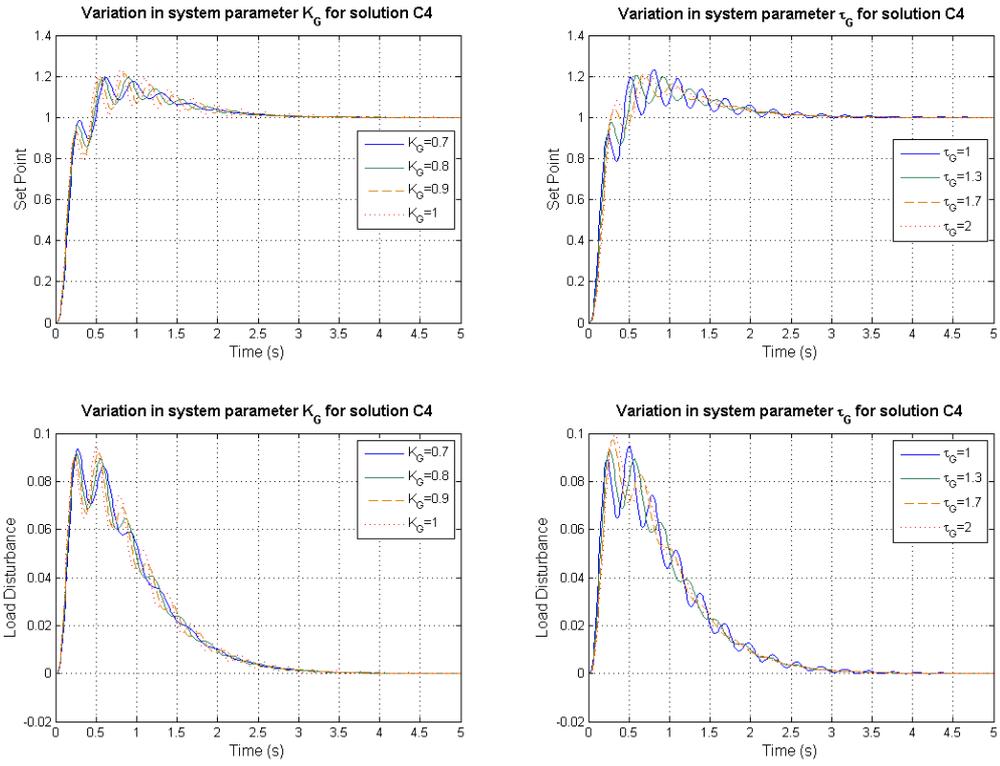

**Figure 13: Robustness analysis of obtained solution C4 for PID controller**

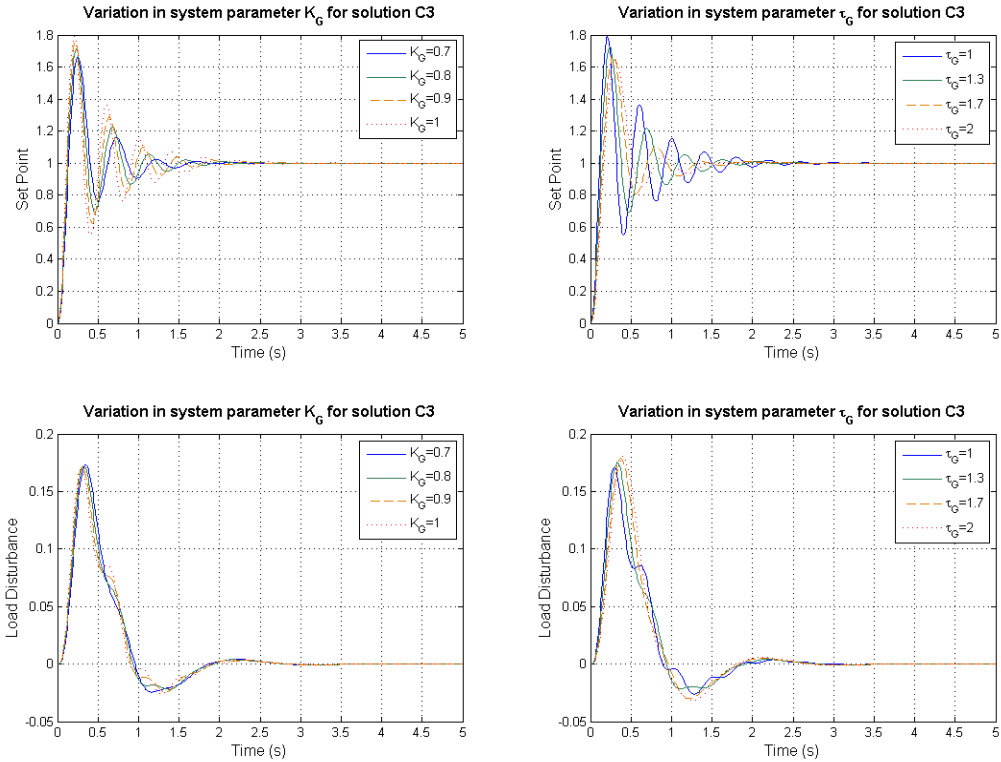

**Figure 14: : Robustness analysis of obtained solution C3 for FOPID controller**

B3 and B4 are the median solutions on the Pareto front which a designer might choose if he wants to obtain a solution which gives average results with respect to both the contradictory objectives. But of-course the designer can give more importance to one objective than the other depending on the specific application. Then the other solutions should be chosen. It is not possible to show robustness analysis of all the controllers on the Pareto front since these are large in numbers. So the robustness of the solutions at the extreme ends of the Pareto front are shown additionally. These are the solutions A3, A4 and C3, C4 as reported in Fig.5 and Table 2. These serve as representative cases and the robustness of the other solutions which lie in between these cases would have similar characteristics. From Figs. 11-14, it is clear that all the obtained solutions on the Pareto front shown sufficient robustness to parameter variation of the system.

The multi-objective optimisation algorithm returns a set of non-dominated solutions on the Pareto front, which show trade-offs for the different objective functions. This set of solutions give the designer an idea of what he can expect out of the controller with respect to different performance indices. When making the final choice for the controller he has to choose one controller from this set depending on which objective function is more important in the design problem. Thus in a particular case for example, let's say that fast tracking is very essential and it is known that the system would need to tolerate less load disturbance. In such a case, the designer can choose a solution on the Pareto front which gives much better tracking than load disturbance. This in effect is actually assigning more importance (weightage) to the objective of set-point tracking than the load disturbance. Though the designer does not physically assign a weight to each of the objective function at the beginning of the optimisation, unlike that practiced in single objective optimization, he chooses a solution looking at the relative trade-off between the objective functions after the completion of the multi-objective optimisation process.

## 9. Conclusions

In this paper, a multi-objective optimization framework is proposed to compare the PID and the FOPID controller for AVR systems. It is shown that none of the controllers are superior than its counterpart for all possible design specifications. For the contradictory objectives of set point tracking and load disturbance rejection, the FOPID controller is better than the PID. On the other hand for set point tracking and lower control signal, the PID outperforms the FOPID. Hence if the control cost is expensive in the design problem, then the PID controller should be preferred. When all three objectives are simultaneously considered, an intersecting Pareto region is found. Hence the PID would be better in some circumstances and the FOPID would be better in others and this would depend on the importance that the designer assigns to the individual objectives on the final Pareto front. Future scope of research can be directed towards frequency domain robust FO controller design to handle uncertainty of the AVR system.

## Appendix

**Table 4: Additional representative solutions on the Pareto fronts for Case I**

| Controller | $J_1$ | $J_2$ | $K_p$ | $K_i$ | $K_d$ | $\lambda$ | $\mu$ |
|---|---|---|---|---|---|---|---|
| FOPID | 1.029145 | 1.044629 | 1.930455 | 0.74026 | 0.429457 | 0.723937 | 1.049256 |
| | 1.033774 | 1.034716 | 1.683444 | 0.94099 | 0.401954 | 0.564563 | 1.041433 |
| | 1.045385 | 1.018859 | 1.266586 | 0.774625 | 0.292459 | 0.575282 | 1.041165 |
| | 1.055533 | 1.011061 | 0.962082 | 0.899883 | 0.312162 | 0.563756 | 1.034014 |
| | 1.065062 | 1.007815 | 0.812938 | 0.783643 | 0.295295 | 0.561241 | 1.03076 |
| | 1.091685 | 1.002933 | 0.530705 | 0.224201 | 0.096227 | 0.890612 | 0.064699 |
| | 1.115204 | 1.002159 | 0.417089 | 0.648128 | 0.253383 | 0.526956 | 1.022238 |
| | 1.131359 | 1.001538 | 0.384709 | 0.295672 | 0.128318 | 0.8501 | 1.048671 |
| | 1.154184 | 1.001138 | 0.288515 | 0.656956 | 0.252714 | 0.518594 | 1.031633 |
| | 1.167854 | 1.00081 | 0.247268 | 0.529831 | 0.162047 | 0.520486 | 1.01843 |
| PID | 1.020161 | 1.096507 | 2.76277 | 0.489925 | 0.48478 | -- | -- |
| | 1.021604 | 1.07448 | 2.4574 | 0.406304 | 0.460698 | -- | -- |
| | 1.024196 | 1.054331 | 2.13427 | 0.432722 | 0.379741 | -- | -- |
| | 1.027283 | 1.042343 | 1.899786 | 0.489151 | 0.379561 | -- | -- |
| | 1.0313 | 1.026978 | 1.540076 | 0.31 | 0.309524 | -- | -- |
| | 1.051236 | 1.011311 | 1.021725 | 0.257186 | 0.148394 | -- | -- |
| | 1.067868 | 1.004338 | 0.642998 | 0.23019 | 0.104529 | -- | -- |
| | 1.100506 | 1.001688 | 0.405165 | 0.194593 | 0.072729 | -- | -- |
| | 1.129208 | 1.000965 | 0.307315 | 0.176972 | 0.037981 | -- | -- |
| | 1.170885 | 1.000569 | 0.236312 | 0.133755 | 0.006267 | -- | -- |

**Table 5: Additional representative solutions on the Pareto front for Case II**

| Controller | $J_1$ | $J_3$ | $K_p$ | $K_i$ | $K_d$ | $\lambda$ | $\mu$ |
|---|---|---|---|---|---|---|---|
| FOPID | 1.008809 | 1.396873 | 1.01585 | 1.339921 | 0.622266 | 0.886173 | 0.999999 |
|  | 1.00887 | 1.352929 | 1.02055 | 1.441694 | 0.627774 | 0.883628 | 0.999999 |
|  | 1.00894 | 1.321245 | 1.017675 | 1.535393 | 0.631745 | 0.880141 | 0.999999 |
|  | 1.009094 | 1.265356 | 1.07598 | 1.700232 | 0.658382 | 0.876623 | 0.999999 |
|  | 1.009321 | 1.219958 | 1.076808 | 1.907612 | 0.665921 | 0.881717 | 0.999999 |
|  | 1.009671 | 1.161284 | 1.150127 | 2.304853 | 0.7037 | 0.854266 | 0.999999 |
|  | 1.010215 | 1.110699 | 1.138668 | 2.982153 | 0.742119 | 0.833934 | 0.999999 |
|  | 1.010962 | 1.073907 | 1.228768 | 3.709496 | 0.769155 | 0.845063 | 0.999999 |
|  | 1.013038 | 1.029954 | 1.392977 | 6.446186 | 0.908295 | 0.844397 | 0.999999 |
|  | 1.039327 | 1.004111 | 4.184707 | 14.46761 | 1.093412 | 0.999999 | 0.999999 |
| PID | 1.013984 | 1.798465 | 6.430902 | 0.518394 | 1.137894 | -- | -- |
|  | 1.014008 | 1.745338 | 6.470968 | 0.541836 | 1.138535 | -- | -- |
|  | 1.014096 | 1.669406 | 6.463039 | 0.581576 | 1.135854 | -- | -- |
|  | 1.014376 | 1.552924 | 6.502451 | 0.654414 | 1.138951 | -- | -- |
|  | 1.014711 | 1.471669 | 6.495779 | 0.719769 | 1.141097 | -- | -- |
|  | 1.015515 | 1.335366 | 6.53658 | 0.872581 | 1.137222 | -- | -- |
|  | 1.016499 | 1.233459 | 6.515949 | 1.061944 | 1.138755 | -- | -- |
|  | 1.017063 | 1.178478 | 6.666353 | 1.221558 | 1.137657 | -- | -- |
|  | 1.018189 | 1.116273 | 6.715426 | 1.529737 | 1.167178 | -- | -- |
|  | 1.019724 | 1.056098 | 6.803907 | 2.244336 | 1.176469 | -- | -- |

**Table 6: Additional representative solutions on the Pareto front for Case III**

| Controller | $J_1$ | $J_3$ | $J_2$ | $K_p$ | $K_i$ | $K_d$ | $\lambda$ | $\mu$ |
|---|---|---|---|---|---|---|---|---|
| FOPID | 1.01929 | 2.67616 | 2.67268 | 0.10829 | 0.48792 | 0.00922 | 0.73879 | 0.15223 |
|  | 1.02480 | 1.96292 | 2.25837 | 0.38446 | 0.70902 | 0.37151 | 0.84631 | 0.60944 |
|  | 1.03199 | 1.95442 | 1.62959 | 0.19993 | 0.59320 | 0.21201 | 0.80003 | 0.18159 |
|  | 1.06060 | 1.26473 | 2.53539 | 0.24284 | 0.57836 | 0.19772 | 0.83591 | 0.18194 |
|  | 1.08648 | 1.21838 | 1.85047 | 0.18770 | 0.59635 | 0.21029 | 0.79219 | 0.16077 |
|  | 1.13943 | 3.34601 | 1.02550 | 0.54682 | 0.99249 | 0.50780 | 0.86723 | 0.68563 |
|  | 1.21656 | 3.64229 | 1.01158 | 0.34135 | 0.60540 | 0.35310 | 0.85007 | 0.39433 |
|  | 1.34341 | 3.68716 | 1.00677 | 0.22045 | 0.61409 | 0.25464 | 0.80328 | 0.27141 |
|  | 1.42517 | 3.95297 | 1.00598 | 0.82796 | 1.50167 | 0.67302 | 0.90004 | 0.61820 |
|  | 1.54743 | 2.24526 | 1.02317 | 0.97308 | 1.67902 | 0.35219 | 0.94531 | 0.68650 |
| PID | 1.02436 | 1.00377 | 2.18047 | 7.36699 | 10.40949 | 1.26046 | -- | -- |
|  | 1.02880 | 1.00612 | 1.77457 | 6.33191 | 8.23201 | 1.32818 | -- | -- |
|  | 1.44059 | 5.10043 | 1.00005 | 0.07139 | 0.09083 | 0.06923 | -- | -- |
|  | 1.96276 | 2.08648 | 1.00265 | 0.48751 | 1.59224 | 0.49996 | -- | -- |
|  | 2.54685 | 2.62786 | 1.00170 | 0.39109 | 1.68470 | 0.47364 | -- | -- |
|  | 3.11255 | 3.35467 | 1.00122 | 0.32945 | 1.52482 | 0.49901 | -- | -- |

|   |   |   |   |   |   |   |   |
|---|---|---|---|---|---|---|---|
| 3.67429 | 3.39173 | 1.00118 | 0.32125 | 1.86738 | 0.49251 | -- | -- |
| 4.13918 | 4.36966 | 1.00086 | 0.27403 | 1.64369 | 0.44773 | -- | -- |
| 4.60375 | 4.75886 | 1.00073 | 0.25265 | 1.51754 | 0.54359 | -- | -- |
| 4.96509 | 4.86126 | 1.00071 | 0.24668 | 1.61100 | 0.53667 | -- | -- |